\documentclass[11pt,fleqn]{article}
\usepackage{graphicx}
\usepackage[a4paper,left=2cm,right=2cm]{geometry}
\usepackage{amsmath}
\usepackage{float}
\usepackage[latin1]{inputenc}
\usepackage{amsmath}
\usepackage{amsfonts}
\usepackage{amssymb}
\usepackage{float}
\floatstyle{boxed} 
\restylefloat{figure}

\newcommand{\be}{\begin{equation}}
\newcommand{\ee}{\end{equation}}
\newcommand{\bea}{\begin{eqnarray}}
\newcommand{\eea}{\end{eqnarray}}

\begin{document}
\title{Search for more sensitive observables to charged scalar in $B \rightarrow D^{(*)}\tau\nu_{\tau}$ decays.}
\author{Lobsang Dhargyal \\\\\ Institute of Mathematical Sciences, Chennai 600113
, India.}
\date{31 March 2017}

\maketitle
\begin{abstract}
It has been known that $B \rightarrow D^{(*)} \tau \nu_{\tau}$ are good observables in the search for the charged Higgs. The recent obervation of deviation from standard-model by almost 4$\sigma$ by Babar, Belle and LHCb in $R(D^{(*)})$ revived the interest in possible signal of presence of charged Higgs in these modes. But such a large deviation in the rates, where standard-model has tree level contribution, coming from a charged Higgs alone is highly unlikely. However these decay modes are good probes to search for small charged Higgs signal if we can construct sensitive observables in these modes. In this work we would like to propose four new observables which shows much more sensitivity to the presence of charged Higgs than the usual observables such as $A_{\lambda}^{D^{(*)}}$ and $A_{\theta}^{D^{(*)}}$. These four observable are (1) $\frac{1}{A_{\lambda}^{D}}$,\ (2) $Y_{1}(q^{2}) = \frac{A^{D}_{\theta}}{A^{D}_{\lambda}}$,\ (3) $Y_{2}(q^{2}) = \frac{d\Gamma(B \rightarrow D^{*}\tau\nu_{\tau})}{d\Gamma_{D}(\lambda_{\tau}=+1/2) - d\Gamma_{D}(\lambda_{\tau}=-1/2)}$ and (4) $Y_{3}(q^{2}) = (\frac{q^{2}}{m^{2}_{\tau}})(A^{D}_{\lambda} + 1)\frac{1}{A^{D}_{\lambda}}$.
\end{abstract}

\section{\large Introduction.}

The LHC discovery of a scalar behaving like the standard-model (SM) Higgs boson \cite{disHig} marks the tentative experimental completion of SM with all the particles it predicted observed experimentally. But even after the LHC discovery of SM like Higgs, still its clear that it is not complete because in SM there is no explanation of Dark Matter and Dark Energy, CP violation due to KM weak phase is turn out to be too small to account for the observed baryon asymmetry of the universe, then there is the strong CP problem and also the fine tunning problem in renormalization of Higgs mass etc. Hence it is pretty evident that we require new-physics (NP) at some scale above about 200 GeV. The absence of clear cut NP signal from both flavor and collider experiments till date may indicate that the scale of NP is much higher than the electro-weak scale. However there are many loophole for low mass NP in current direct search by LHC due to sensitivity limits of LHC to light weakly coupled particles. But there has been reported many 2-4 $\sigma$ deviations in B meson decays by BABAR, Belle and LHCb recently, some of which could the tip of the iceberg signals of NP. The reported deviations from SM predictions by Babar \cite{Babar} and Belle \cite{Belle1}\cite{Belle}\cite{Bellelatest} in $R(D^{(*)}) = \frac{Br(B \rightarrow D^{(*)}\tau\nu)}{Br(B \rightarrow D^{(*)}l\nu)}$ and also LHCb \cite{LHCb} has reported an excess in $R(D^{*})$ consistent with Babar and Belle results is the strongest hints of a possible lepton flavor universality violating NP in b quark and/or $\tau$ lepton sector. Here $l$ refers to $e$ or $\mu$. The present world average from heavy-flavor-averaging-group (HFAG) of these measurements is \cite{Greg34}
\be
\begin{split}
R(D)_{EXP} = 0.397 \pm 0.040 \pm 0.028\\
R(D^{*})_{EXP} = 0.316 \pm 0.016 \pm 0.010.
\end{split}
\label{Eq:1}
\ee
Comparing these measurement with the SM predictions \cite{Greg17}\cite{Greg18}
\be
\begin{split}
R(D)_{SM} = 0.300 \pm 0.008\\
R(D^{*})_{SM} = 0.252 \pm 0.003,
\end{split}
\ee
there is a deviation of 2 $\sigma$ for the R(D) and 3.4 $\sigma$ for the $R(D^{*})$. Taking the negative correlation of about -0.23 \cite{Acel2} between the two data into account the combine deviation from SM is close to 4$\sigma$. It is further supported by measurement of Br($B \rightarrow \tau \nu$) by Babar\cite{Babar1} and Belle\cite{Belle1} with HFAG average of \cite{Greg30}
\be
Br_{EXP}(B \rightarrow \tau \nu) = (1.06 \pm 0.19)\times 10^{-4},
\ee
which is 1.4 $\sigma$ above the SM prediction \cite{Greg14}
\be
Br_{EXP}(B \rightarrow \tau \nu) = (0.75 \pm 0.1)\times 10^{-4}.
\ee
Babar \cite{Babar} and Belle \cite{Belle} have ruled out 2HDM type-II at 99.8\% CL from disagreement of its prediction with data as an explanation of the anomalies in $R(D)$ and $R(D^{*})$. From the on set it is very easy to see that this anomalies can be explained by a non universal left handed vector particle but a simple non-universally interacting heavier gauge boson ($W^{'\pm}$) is highly constrained by null results from LHC search for $W^{'} \rightarrow t\bar{b}$ signals \cite{Greg35}\cite{Greg36}, and also by precision measurements in $\mu$ \cite{Greg37} and $\tau$ \cite{Greg38}. Therefore as of now it is very difficult to built a non-universal gauge model that can fit all the constrains and so in this paper we will mostly strick to a model-independent analysis only. It has been shown first in references \cite{Fajfer}\cite{T-W}\cite{Ysaki}\footnote{as far as author knows} that the observed excess in $R(D^{(*)})$ can be explained with baryon and lepton number conserving Lepto-quark (LQ) models and followed in with many special cases and variations of the LQ models has been proposed to explain not only $R(D^{(*)})$ but also observed deviations in $R_{K} = \frac{Br(B \rightarrow K \mu\mu)}{Br(B \rightarrow K e e)}$ and the so called $P^{'}_{5}$ anomalies. But some of these LQ models turn out to be not viable when all precision data till date are taken into account, for details see the recent review in \cite{Diego}. In any case as of now the experimental inputs seems to be too few and far apart to build a complete and consistent NP model if at all NP shows up at the reach of the upgraded LHC and Belle-II. In following sections we will give a general model-independent analysis of possible contribution from charged scalar to these deviations and observables sensitive to their presence. This paper is organized as follows: In Section II we present the general formulism of the analysis and lay the theoretical framework of the paper. Section III contains an introduction to observable sensitive to NP. Section IV contains the core of this work and it deals with new and more sensitive observables to the presence of charged scalar NP. In section V we conclude the paper.

\section{Theoretical Framework.}

We assume that all the neutrinos is are left handed, then the most general effective Hamiltonian that contains all possible four-fermion operators of dimention four for the decay process $b \rightarrow cl\nu_{l}$, where $l = \tau$, $\mu$ or $e$ here, is given as \cite{Ysaki}
\be
\mathcal{H}_{eff} = \frac{4G_{F}}{\sqrt{2}}V_{cb}[ (\delta_{ll} + C_{V_{L}}^{l})\mathcal{O}_{L}^{l} + C_{V_{R}}^{l}\mathcal{O}_{R}^{l} - C_{S_{L}}^{l}\mathcal{O}_{S_{L}}^{l} - C_{S_{R}}^{l}\mathcal{O}_{S_{R}}^{l} + C_{T}^{l}\mathcal{O}_{T}^{l} ]
\label{Eq:5}
\ee
with the operators define as\\
\\
$\mathcal{O}_{L}^{l} = (\bar{c}_{L}\gamma^{\mu}b_{L})(\bar{l}_{L}\gamma_{\mu}\nu_{lL})$,\ $\mathcal{O}_{R}^{l} = (\bar{c}_{R}\gamma^{\mu}b_{R})(\bar{l}_{L}\gamma_{\mu}\nu_{lL})$,\\
\\
$\mathcal{O}_{S_{L}}^{l} = (\bar{c}_{R}b_{L})(\bar{l}_{R}\nu_{lL})$,\ $\mathcal{O}_{S_{R}}^{l} = (\bar{c}_{L}b_{R})(\bar{l}_{R}\nu_{lL})$ and
\be
\mathcal{O}_{T}^{l} = (\bar{c}_{R}\sigma^{\mu\nu}b_{R})(\bar{l}_{L}\gamma_{\mu\nu}\nu_{lL}).
\ee
In Eqs.(\ref{Eq:5}) we have explicitly shown the relative negative sign between effective four fermion operators due to exchange of heavy scalar particles and heavy vector particles. This is due to sign difference between a scalar propagator and a vector propagator\footnote{This is why in forces mediated by exchange of scalars, particles carrying same charges attract towards each other while in forces mediated by exchange of vector particles, particles with same charges repel each other.}. In many analysis the relative sign is implicitly absorbed into the effective coefficients, but if the relative sign between the vector four current operators and the scalar four current operators are explicitly shown will help us rule out few models, where NP is scalar and real parts of $C_{S_{L}}^{l}$ and $C_{S_{R}}^{l}$ are dominant, given that we expect NP contribution is less than the SM contribution. Fore instance, in 2HDM of type-I and type-II, the effective coupling are real and positive and so these type of models will interferes destructively with SM, due to the relative negative sign, and so 2HDM of type-I and type-II can only reduce the values of $R(D^{(*)})$ instead of increasing it as required by experiments in all the parameter spaces where the NP part is less than SM part. So it is clear from this that the relative sign can actually help us in ruling out all the models of new scalar particles whose effective coupling are non-negative for the most parts of the parameter space where NP part is less than the SM part. In this work we will not deal with new vector and tensor terms. So the following analysis is important if in the future experiments in these modes, presence of only scalar type NP is found, then we need new and more sensitive observables to better differentiate the NP from SM. In the next section we give a brief analysis on how future experiments in these modes can differentiate the presence of scalar NP from vector NP, scalar NP from tensor type NP and scalar NP from the presence of both vector and tensor type NP. Now then with presence of only scalar and vector (SM) type operators remaining we can express the effective Hamiltonian in Eqs.(\ref{Eq:5}) as
\be
\mathcal{H}_{eff} = \frac{G_{F}}{\sqrt{2}}V_{cb}[ (\bar{c}\gamma^{\mu}(1 - \gamma_{5})b)(\bar{l}\gamma_{\mu}(1 - \gamma_{5})\nu_{l}) - (\bar{c}(\epsilon_{S_{l}} + \epsilon_{P_{l}}\gamma_{5})b)(\bar{l}(1 - \gamma_{5})\nu_{l}) ]
\label{Eq:eff}
\ee
where
\be
\epsilon_{s_{l}} = C_{S_{R}}^{l} + C_{S_{L}}^{l},\ \epsilon_{p_{l}} = C_{S_{R}}^{l} - C_{S_{L}}^{l}.
\label{Eq:simp-coupling}
\ee
The most stringent B physics constrains on the scalar NP explanation of $R(D^{(*)})$ comes from the decay rates $Br(B_{c} \rightarrow \tau\nu_{\tau})$ or $Br(B_{u} \rightarrow \tau\nu_{\tau})$ depending on the particularities of the NP model. So in what follows we will take these observables and their measured bounds as additional constrains, wherever applicable, when fixing the coefficients of the effective operators to $R(D^{(*)})$ data. Assuming all hadronization are due to strong interaction, due to parity conservation of strong force, only scalar and vector current can contribute in $R(D)$ and so it only constrains the $\epsilon_{s_{l}}$ and in the case of $Br(B_{c} \rightarrow \tau\nu_{\tau})$ and $Br(B_{u} \rightarrow \tau\nu_{\tau})$, only pseudo-scalar and axial-vector current can contribute and so these observables only constrain $\epsilon_{p_{l}}$. However to $R(D^{*})$, both vector and axial-vector currents can contribute but only pseudo-scalar current can contribute and so $R(D^{*})$ constrains $\epsilon_{p_{l}}$. In presence of charged scalar particle, the differential decay rate of $B \rightarrow D^{(*)}\tau\nu_{\tau}$ can be expressed as \cite{Jung}
\be
\begin{split}
\frac{d\Gamma(B \rightarrow D\tau\nu_{\tau})}{dq^{2}} = \frac{G_{F}^{2}|V_{cb}|^{2}|\vec{p}_{D}|q^{2}}{96\pi^{3}m_{B}^{2}}(1-\frac{m_{\tau}^{2}}{q^{2}})^{2}\{|H_{0}|^{2}(1+\frac{m_{\tau}^{2}}{q^{2}})\\
+ \frac{3m_{\tau}^{2}}{2q^{2}}|H_{t}|^{2}[(1 - \frac{q^{2}}{m_{\tau}(m_{b}-m_{c})}Re(\epsilon^{\tau}_{s}))^{2} + \frac{q^{4}}{m_{\tau}^{2}(m_{b}-m_{c})^{2}}(Im(\epsilon^{\tau}_{s}))^{2}]\}
\label{Eq:dT1}
\end{split}
\ee
and
\be
\begin{split}
\frac{d\Gamma(B \rightarrow D^{*}\tau\nu_{\tau})}{dq^{2}} = \frac{G_{F}^{2}|V_{cb}|^{2}|\vec{p}_{D^{*}}|q^{2}}{96\pi^{3}m_{B}^{2}}(1-\frac{m_{\tau}^{2}}{q^{2}})^{2}\{(|H_{00}|^{2} + |H_{--}|^{2} + |H_{++}|^{2})(1+\frac{m_{\tau}^{2}}{q^{2}})\\
+ \frac{3m_{\tau}^{2}}{2q^{2}}|H_{0t}|^{2}[(1 - \frac{q^{2}}{m_{\tau}(m_{b}-m_{c})}Re(\epsilon^{\tau}_{p}))^{2} + \frac{q^{4}}{m_{\tau}^{2}(m_{b}-m_{c})^{2}}(Im(\epsilon^{\tau}_{p}))^{2}]\}.
\end{split}
\label{Eq:dT2}
\ee
Now from Eqs.(\ref{Eq:dT1}) and Eqs.(\ref{Eq:dT2}) we can see that models where $Re(\epsilon^{\tau}_{/sp})$, $Im(\epsilon^{\tau}_{s/p}) <$ 1, the dominant contribution comes from the $Re(\epsilon^{\tau}_{s/p})$ as it has a term linear in $Re(\epsilon^{\tau}_{s/p})$ from the mixing with the SM part where as $Im(\epsilon^{\tau}_{s/p})$ enters only in quadratic powers. As seen from the above two equations, the relative sign does not effect the contribution from the complex part of new physics but it affects the contributions from the real part of new scalars. In any case, whether the $\epsilon^{\tau}_{s/p}$ are real or complex, the new observables that we will introduce in the following sections are more sensitive towards presence of scalar NP then the previously existing observables. For details of relation between vector, axial-vector, scalar, psuedo-scalar and tensor currents and their respective form factors see \cite{T-W}\cite{Ysaki}\cite{Acel1}. For numerical values of the parameters in the form factors, we will use those given in \cite{Ysaki} with exception that we will use $R_{3}(1) = 0.97$ instead of $R_{3}(1) = 1.22$ of that reference.

\section{Observables sensitive to NP.}

With lack of any persistent sign of NP from direct searches at LHC, the precision physics is becoming more and more important to at-least sense the direction of the possible nature of NP. So it has become crucial to find sensitive observables to NP that can be tested in flavor precision machines such as Belle II and LHCb etc. The remaining part of this work is concern with finding more sensitive observables than the usual ones like tau spin asymmetry, $A_{\lambda}^{D^{(*)}}$, and forward-backward asymmetries, $A_{\theta}^{D^{(*)}}$, which will be defined in the following sections. We will be mainly concerned with charged scalar NP and define four very sensitive new observables to charged scalar NP in this work.

\subsection{Observables sensitive to non-scalar NP.}
\label{sec:3}

In case of new vector particles with substantial couplings to vector and axial-vector currents, since only vector current will contribute to hadronization in $R(D)$, $R(D)$ constrains only the vector coupling (1+$\epsilon_{v_{NP}}$). Where we will denote by $\epsilon_{v_{NP}}$ and $\epsilon_{a_{NP}}$, the effective couplings of new vector particles to vector and axial-vector effective four currents respectively. Now since $R(D) = \frac{Br(B \rightarrow D\tau\nu)}{Br(B \rightarrow Dl\nu)}$, we have
\be 
\frac{R(D)_{NP}}{R(D)_{SM}} = |1 + \epsilon_{v_{NP}}|^{2} = \frac{0.397}{0.300} = 1.323,
\ee
and for real $\epsilon_{v_{NP}}$ and $\epsilon_{a_{NP}}$, we have $\epsilon_{v_{NP}} = 0.150$ gives $R(D) = 0.397$. Now using this value of $\epsilon_{v_{NP}}$ in $R(D^{*})$ which gets contributions from both vector current and axial-vector current, we can fit the $R(D^{*}) = 0.316$ if we set $\epsilon_{a_{NP}} = 0.121$. So a new vector particle which couples to the different generations of fermions differently can explain the observed excess in the $R(D^{(*)})$ easily. Now if the observed excess in $R(D^{(*)})$ has some contribution due to new vector particles, then as pointed out in \cite{Acel1}, the observable
\be
X_{1}(q^{2}) = R(D^{*}) - R(D^{*}_{L})
\ee
is independent from effects due to presence of any new scalar particles, and so this observable also should show excess similar to $R(D^{*})$, where $R(D^{*}_{L})$ refers to the ratio for the longitudinally polarized $D^{*}$. 
Another observable which can be used to check the presence of new non-scalar particles contributing to $R({D^{(*)}})$ are define as \cite{Ysaki}\cite{Acel1}
\be
X_{2}^{D}(q^{2}) = R_{D}(q^{2})(A^{D}_{\lambda} - 1)\ and\ X_{2}^{D^{*}}(q^{2}) = R_{D^{*}}(q^{2})(A^{D^{*}}_{\lambda} -1)
\ee
where the $A^{D}_{\lambda}$ and $A^{D^{*}}_{\lambda}$ are the $\tau$ spin-asymmetry defined as \cite{Fajfer}\cite{Adatta}\cite{Mtanaka}\cite{Mtanaka2}
\be
A^{D}_{\lambda} = \frac{d\Gamma^{D}(\lambda = +1/2)dq^{2} - d\Gamma^{D}(\lambda = -1/2)dq^{2}}{d\Gamma^{D}(\lambda = +1/2)dq^{2} + d\Gamma^{D}(\lambda = -1/2)dq^{2}} = \frac{\frac{3m_{\tau}^{2}}{2q^{2}}|H_{t}|^{2} - (1 - \frac{m_{\tau}^{2}}{2q^{2}})|H_{0}|^{2}}{\frac{3m_{\tau}^{2}}{2q^{2}}|H_{t}|^{2} + (1 + \frac{m_{\tau}^{2}}{2q^{2}})|H_{0}|^{2}}
\label{Eq:TspinAym-D}
\ee
and
\be
\begin{split}
A^{D^{*}}_{\lambda} = \frac{d\Gamma^{D^{*}}(\lambda = +1/2)dq^{2} - d\Gamma^{D^{*}}(\lambda = -1/2)dq^{2}}{d\Gamma^{D^{*}}(\lambda = +1/2)dq^{2} + d\Gamma^{D^{*}}(\lambda = -1/2)dq^{2}} = \frac{\frac{3m_{\tau}^{2}}{2q^{2}}|H_{0t}|^{2} - (1 - \frac{m_{\tau}^{2}}{2q^{2}})[|H_{00}|^{2} + |H_{--}|^{2} + |H_{++}|^{2}]}{\frac{3m_{\tau}^{2}}{2q^{2}}|H_{0t}|^{2} + (1 + \frac{m_{\tau}^{2}}{2q^{2}})[|H_{00}|^{2} + |H_{--}|^{2} + |H_{++}|^{2}]}.
\end{split}
\label{Eq:TspinAym-Dstr}
\ee
As shown in a general analysis in presence of new scalar, vector and tensor type operators in \cite{Ysaki}, the observables $X_{2}^{D^{(*)}}(q^{2})$ are independent of contributions from the scalar NP. So in future measurements in these modes, if the deviations in $R(D^{(*)})$ is remains and a comparble deviations in $X_{2}^{D^{(*)}}(q^{2})$ are found, then we can be sure that the most dominant NP is a non-scalar NP.  Now from Eqs.(\ref{Eq:TspinAym-Dstr}) we can see that, since $H_{0t}$ and $H_{00}$ depends only on axial vector and psuedo-scalar current form factors, if $R(D^{*})$ shows deviation from SM but $A^{D^{*}_{L}}_{\lambda}$ is consistent with SM, then the scalar and tensor contribution is negligible and a new vector boson with substantial coupling to the axial-vector current is the most likely NP. Similarly if $R(D)$ shows deviation from SM but $A_{D}^{\lambda}$ does not show any noticeable deviation, then the scalar and tensor contribution is negligible and a new vector boson with substantial coupling to the vector current is the most likely NP. If in future experiments in these modes, we found that $X_{2}^{D^{(*)}}(q^{2})$, $A_{D}^{\lambda}$, $A^{D^{*}_{L}}_{\lambda}$ and $R(D^{(*)})$ all shows comaprable deviations from SM, then we can be sure of presence of tensor type NP or vector and tensor type NP and no or atleast neglegible presence of scalar type NP. In what follows, we will assume the scenario where future measurements in these modes finds that deviations in $R(D^{(*)})$ remains but no comparable deviations are found in the observables $X_{2}^{D^{(*)}}(q^{2})$, a clear sign of presence of scalar type NP. Then we will need the new and more sensitive observables that we propose in the following sections to better probe the presence of scalar NP in these modes.

\subsection{Observables sensitive to charged scalar.}

Besides $R(D)(q^{2})$ and $R(D^{*})(q^{2})$, we can define many more observables that are sensitive to the presence of new charged scalar particles in the $B \rightarrow D^{(*)} \tau \nu_{\tau}$ decay distributions. One such observable is tau spin asymmetry ($A^{D^{(*)}}_{\tau}$) which is already defined in Eqs.(\ref{Eq:TspinAym-D}) and Eqs.(\ref{Eq:TspinAym-Dstr}) of section \ref{sec:3}. Another is the forward-backward asymmetries define as \cite{Fajfer}\cite{Acel1}
\be
A^{D^{(*)}}_{\theta} = \frac{\int^{0}_{-1}d\cos{\theta}(d^{2}\Gamma^{D^{(*)}}_{\tau}/dq^{2}d\cos{\theta}) - \int^{1}_{0}d\cos{\theta}(d^{2}\Gamma^{D^{(*)}}_{\tau}/dq^{2}d\cos{\theta})}{d\Gamma^{D^{(*)}}_{\tau}/dq^{2}}
\ee
which can be expresses as
\be
A^{D}_{\theta} = \frac{3m_{\tau}^{2}}{2q^{2}}\frac{Re(H_{0}\bar{H}^{*}_{t})}{|H_{0}|^{2}(1+\frac{m_{\tau}^{2}}{2q^{2}}) + \frac{3m_{\tau}^{2}}{2q^{2}}|\bar{H}_{t}|^{2}}
\ee
and
\be
A^{D^{*}}_{\theta} = \frac{3}{4}\frac{[|H_{++}|^{2} - |H_{--}|^{2} + 2\frac{m_{\tau}^{2}}{q^{2}}Re(H_{00}\bar{H}^{*}_{0t})]}{[(|H_{--}|^{2} + |H_{++}|^{2} + |H_{0}|^{2})(1+\frac{m_{\tau}^{2}}{2q^{2}}) + \frac{3m_{\tau}^{2}}{2q^{2}}|H_{t}|^{2}]}
\ee
where the bar over $H_{t}$ and $H_{0t}$ refers to $H_{t}(1 - \frac{q^{2}}{m_{\tau}(m_{b}-m_{c})}\epsilon_{s})$ and $H_{0t}(1 - \frac{q^{2}}{m_{\tau}(m_{b}+m_{c})}\epsilon_{p})$ respectively. The forward-backward asymmetry is important because $R(D^{(*)})(q^{2})$, $R(D^{*}_{L})(q^{2})$ and $A^{D^{(*)}}_{\tau}$ do not give independent information, as they can be expressed in terms of each other using $X_{1}^{D^{(*)}}$ and $X_{2}^{D^{(*)}}$, and so only $A^{D^{(*)}}_{\theta}$ are independent constrains in the complex $\epsilon$ planes \cite{Acel1}.

\section{More sensitive observables to charged scalar.}

The question is, $A^{D^{(*)}}_{\tau}$ and $A^{D^{(*)}}_{\theta}$ are observables sensitive to charged scalars but can we construct new observables which are more sensitive to charged scalars than $A^{D^{(*)}}_{\tau}$ and $A^{D^{(*)}}_{\theta}$? In what follows we will give an affirmative answer to this question by giving four new observables which are more sensitive to the presence of charged scalars than $A^{D^{(*)}}_{\tau}$ and $A^{D^{(*)}}_{\theta}$. To show sensitivity of new observables, we will use the scalar parameters from the model given in \cite{our1}\cite{our2}, where the effective scalar couplings ($\epsilon$) are same as in Type-II 2HDM for the $\mu$ and e but enhenced by a factor $\eta$ in the $\tau$ sector. This model corresponds to the $\eta = -1$ case in a 2HDM of type-II, where $\eta$ is a anomalous multiplicative factor only affecting the $\tau$ or b Yukawa coupling with charged Higgs, given in the comments at the end of reference \cite{our2} about the anomalous SUSY. In that model $\epsilon_{s}^{\tau} \approx \epsilon_{p}^{\tau} \approx \epsilon^{\tau} = -m_{b}m_{\tau}\frac{\tan{\beta}^{2}}{M_{H^{\pm}}^{2}}$ where as $\epsilon_{s}^{e,\mu} \approx \epsilon_{p}^{e,\mu} \approx \epsilon^{e,\mu} = +m_{b}m_{e,\mu}\frac{\tan{\beta}^{2}}{M_{H^{\pm}}^{2}}$. The most important constrains in this model comes from the $Br(B_{c} \rightarrow \tau \nu_{\tau})$ and $Br(B_{u} \rightarrow \tau \nu_{\tau})$ in fitting $R(D^{(*)})$. But since $Br(B_{c} \rightarrow \tau \nu_{\tau})$ is not measured yet and theoretical estimations still allows $Br(B_{c} \rightarrow \tau \nu_{\tau})$ from $5\%$ to $30\%$ \cite{Alonso} compare to SM value of $2.22\%$ \cite{Xinli}. So as of now $Br(B_{u} \rightarrow \tau \nu_{\tau}) = (1.06 \pm 0.19)\times 10^{-4}$, which is 1.4$\sigma$ in excess of SM value, is the most important constrain on scalar parameters in fitting $R(D^{(*)})$. Now from fitting $R(D^{(*)})$ and $Br(B \rightarrow \tau \nu_{\tau})$ simultaneously we get the best fit value of $\frac{\tan\beta}{M_{H^{\pm}}}$ as $\frac{\tan{\beta}}{M_{H^{\pm}}} = 0.098 \pm 0.020$, which gives the lepton ($l$) mass independent charge scalar parameter contributing to $\bar{H}_{t}$ and $\bar{H}_{0t}$, see Eqs.(\ref{Eq:dT1}) and Eqs.(\ref{Eq:dT2}), as
\be
\frac{\epsilon^{l}_{s}}{m_{l}} = \frac{\epsilon^{l}_{p}}{m_{l}} = \frac{\epsilon^{l}}{m_{l}} = \mp m_{b}(\frac{\tan\beta}{M_{H^{\pm}}})^{2} = \mp 0.041 \pm 0.016
\label{Eq:epsilon}
\ee
where the upper sign is for the $\tau$ lepton and lower sign is for the e and $\mu$ leptons.\\
\\
These values of $\frac{\epsilon^{l}}{m_{l}}$ give
\be
Br(B \rightarrow \tau \nu_{\tau})_{NP} = (1.21 \pm 0.783)\times 10^{-4},
\ee
\be
R(D)_{NP} = 0.340 \pm 0.197
\label{Eq:1-mod1}
\ee
and
\be
R(D^{*})_{NP} = 0.255 \pm 0.067.
\label{Eq:1-mod2}
\ee
Comparing Eqs.(\ref{Eq:1-mod1},\ref{Eq:1-mod2}) to Eqs.(\ref{Eq:1}), we can see that the model prediction fits the combine $R(D^{(*)})$ data within 1$\sigma$ of the experimental values.

\subsection{$\frac{1}{A^{D}_{\lambda}}$}

One of the the most sensitive new observable to charged scalars that we can construct turn out to be $\frac{1}{A^{D}_{\tau}}$ given as
\be
\frac{1}{A^{D}_{\lambda}} = \frac{\frac{3m_{\tau}^{2}}{2q^{2}}|\bar{H}_{t}|^{2} + (1 + \frac{m_{\tau}^{2}}{2q^{2}})|H_{0}|^{2}}{\frac{3m_{\tau}^{2}}{2q^{2}}|\bar{H}_{t}|^{2} - (1 - \frac{m_{\tau}^{2}}{2q^{2}})|H_{0}|^{2}}
\label{Eq:1ByTspinAym-D}.
\ee
\begin{figure}[h!]
\begin{minipage}[t]{0.48\textwidth}
\hspace{0.4cm}
\includegraphics[width=0.9\linewidth, height=6cm]{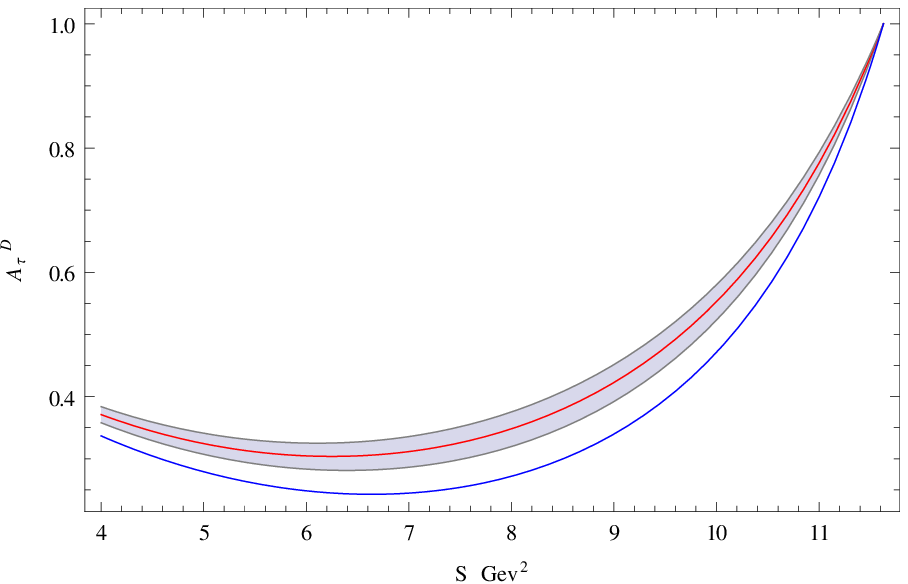}
\end{minipage}
\begin{minipage}[t]{0.48\textwidth}
\hspace{0.4cm}
\includegraphics[width=1\linewidth, height=6cm]{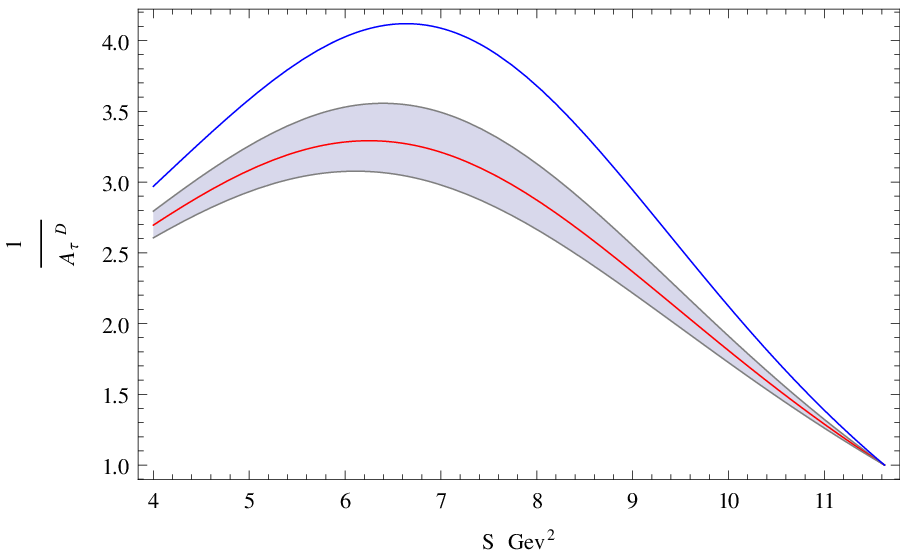}
\end{minipage}
\caption{This figure shows the plots of $A^{D}_{\tau}$ (Left) and $\frac{1}{A^{D^{(*)}}_{\tau}}$ (Right) where the Blue (SM) and the Red (NP) with the 1 $\sigma$ error bar shown around the NP plot. Here $\epsilon^{\tau}_{s} = \epsilon^{\tau}_{p} = -0.072 \pm 0.028$ is used from Eqs.(\ref{Eq:epsilon}).}
\label{Fig:fig1}
\end{figure}
This $\frac{1}{A^{D}_{\tau}}$ observable has two key features that makes it a better observable than $A^{D}_{\tau}$. First as seen from the Figure \ref{Fig:fig1}, although $A^{D}_{\tau}$ shows pretty good sensitivity to NP, but $\frac{1}{A^{D}_{\tau}}$ is more sensitive to NP in the entire range of the plot. Secondly in contrast to $A^{D}_{\tau}$, $\frac{1}{A^{D}_{\tau}}$ show much more sharp maxima, which can be used to measure the shift in the position of the maxima of the SM and the NP due to presence of charged scalar. For the left plot in Figure \ref{Fig:fig1}, the SM (Blue) maxima of $\frac{1}{A^{D}_{\tau}}$ occurs at $q^{2} = S = 6.637$ with maxima value of 4.119 where as the maxima with the presence of scalar NP (Red) occurs at $q^{2} = S = 6.250$ with maxima value of 3.292, where $q^{2} = S = (p_{B}-p_{D})^{2}$ is the momentum transfered squared. Now the difference between the position of the SM maxima and maxima due to scalar NP is 0.387. So if the experimental error in measurements of the position of this maxima and the error in the prediction of the position of the SM maxima can be reduced such that the combined experimental error and theoretical (SM) error can be reduce below 0.078, then we can have a 5$\sigma$ discovery potential for scalar NP with $\epsilon$ as small as -0.072. Now when integrated in $q^{2}$ for the $\frac{1}{A^{D}_{\lambda}}$ we have\\
$\frac{\int^{(m_{B}-m_{D})^{2}}_{m_{\tau}^{2}}dq^{2}d\Gamma /dq^{2}}{\int^{(m_{B}-m_{D})^{2}}_{m_{\tau}^{2}}dq^{2}[d\Gamma(\lambda = +1/2) /dq^{2} - d\Gamma(\lambda = -1/2) /dq^{2}]}|_{SM}$
\be
\begin{split}
- \frac{\int^{(m_{B}-m_{D})^{2}}_{m_{\tau}^{2}}dq^{2}d\Gamma /dq^{2}}{\int^{(m_{B}-m_{D})^{2}}_{m_{\tau}^{2}}dq^{2}[d\Gamma(\lambda = +1/2) /dq^{2} - d\Gamma(\lambda = -1/2) /dq^{2}]}|_{NP} = 0.551
\end{split}
\label{Eq:int-1}
\ee
where as for the $A^{D}_{\lambda}$ we have\\

$\frac{\int^{(m_{B}-m_{D})^{2}}_{m_{\tau}^{2}}dq^{2}[d\Gamma(\lambda = +1/2) /dq^{2} - d\Gamma(\lambda = -1/2) /dq^{2}]}{\int^{(m_{B}-m_{D})^{2}}_{m_{\tau}^{2}}dq^{2}d\Gamma /dq^{2}}|_{SM}$
\be
\begin{split}
- \frac{\int^{(m_{B}-m_{D})^{2}}_{m_{\tau}^{2}}dq^{2}[d\Gamma(\lambda = +1/2) /dq^{2} - d\Gamma(\lambda = -1/2) /dq^{2}]}{\int^{(m_{B}-m_{D})^{2}}_{m_{\tau}^{2}}dq^{2}d\Gamma /dq^{2}}|_{NP} = 0.070.
\end{split}
\label{Eq:int-2}
\ee
From Eqs.(\ref{Eq:int-1}) and Eqs.(\ref{Eq:int-2}) we see that $\frac{1}{A_{\lambda}^{D}}$ is an order of magnitude more sensitive to charged scalar than $A_{\lambda}^{D}$. So in the $q^{2}$ integrated ratios of $A_{\lambda}^{D}$ and $\frac{1}{A_{\lambda}^{D}}$, for $A_{\lambda}^{D}$ we require SM plus Experimental combine error to reduce below 0.014 where as for $\frac{1}{A_{\lambda}^{D}}$ we only require SM plus Experimental error to reduce below 0.110 to have 5$\sigma$ discovery potential of scalar NP with $\epsilon^{\tau}$ as small as -0.072. One may expect that another potential sensitive observable would be $\frac{1}{A^{D^{*}}_{\tau}}$, but due to suppression of scalar NP contribution in $D^{*}$ mode by a factor of $\frac{m_{b} - m_{c}}{m_{b} + m_{c}}$ relative to D mode, the observables $A^{D^{*}}_{\tau}$ and $\frac{1}{A^{D^{*}}_{\tau}}$ are not that sensitive and so we will not use these observables in this work.

\subsection{$Y_{1}(q^{2}) = \frac{A^{D}_{\theta}}{A^{D}_{\lambda}}$}
Another sensitive observable to charged scalar NP can be defined as
\be
Y_{1}(q^{2}) = \frac{A^{D}_{\theta}}{A^{D}_{\lambda}} = \frac{\frac{3m^{2}_{\tau}}{2q^{2}}Re(H_{0}\bar{H}^{*}_{t})}{\frac{3m_{\tau}^{2}}{2q^{2}}|\bar{H}_{t}|^{2} - (1 - \frac{m_{\tau}^{2}}{2q^{2}})|H_{0}|^{2}}\ .
\ee

\begin{figure}[h!]
\begin{minipage}[t]{0.48\textwidth}
\hspace{0.4cm}
\includegraphics[width=0.9\linewidth, height=6cm]{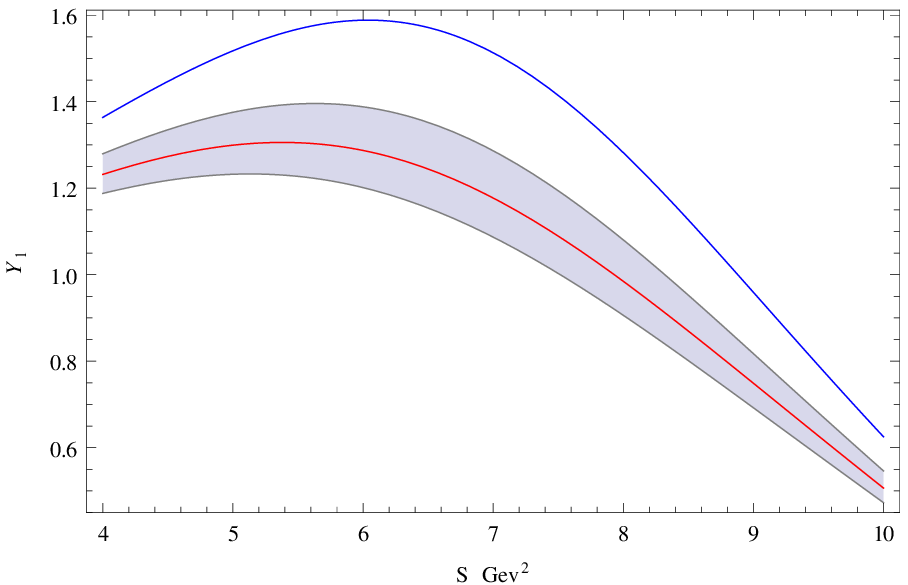}
\end{minipage}
\begin{minipage}[t]{0.48\textwidth}
\hspace{0.4cm}
\includegraphics[width=1\linewidth, height=6cm]{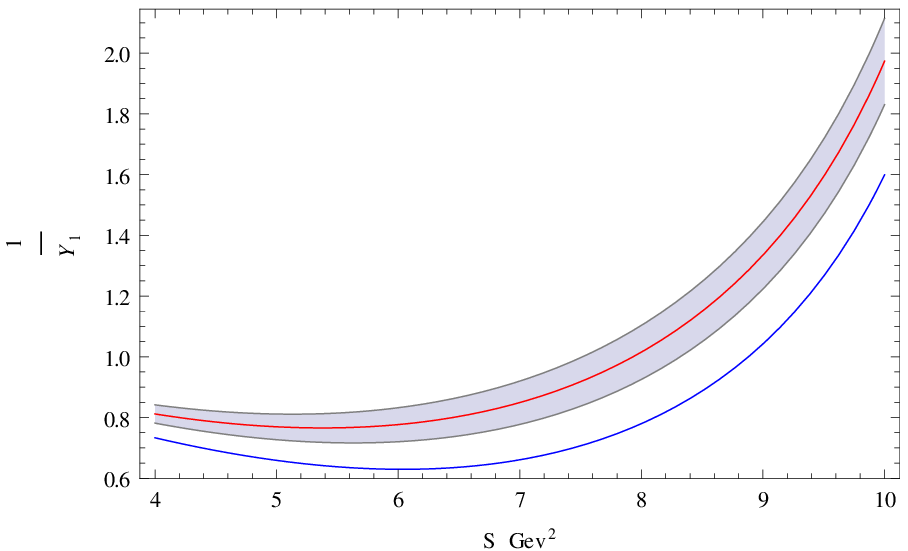}
\end{minipage}
\caption{This figure shows the plots of $Y_{1}(q^{2})$ (left) and $\frac{1}{Y_{1}(q^{2})}$ (Right) where the Blue (SM) and the Red (NP) with the 1 $\sigma$ error bar shown around the NP plot. Here $\epsilon^{\tau}_{s} = \epsilon^{\tau}_{p} = -0.072 \pm 0.028$ is used from Eqs.(\ref{Eq:epsilon}).}
\label{Fig:fig2}
\end{figure}
In Figure \ref{Fig:fig2} we have shown the plot of $Y_{1}(q^{2})$ (left) and $\frac{1}{Y_{1}(q^{2})}$ (Right).\footnote{one may think that actually $\frac{2q^{2}}{3m_{\tau}^{2}}Y_{1}(q^{2})$ will be more sensitive but it turns out that is not the case, in fact the opposite case turn out to be true!} As seen form that Figure, besides showing prominent difference between SM and scalar NP, in the two plots, the $Y_{1}(q^{2})$ is more sensitive towards the low $q^{2}$ region where as the $\frac{1}{Y_{1}(q^{2})}$ is more sensitive towards the high $q^{2}$ region. And one of the most important feature of these observables turn out to be the difference between the position of SM maxima and the scalar NP maxima in $Y_{1}(q^{2})$. The SM (Blue) maxima occurs at $q^{2} = S = 6.038$ with the maxima value of 1.589 where as the scalar NP (Red) maxima occurs at $q^{2} = S = 5.379$ with the maxima value of 1.306, the difference between the positions of the two maximas is 0.659. So if the experimental error in measurements of the position of the maxima and the error in the prediction of the position of the SM maxima can be reduced such that the combined experimental error and theoretical (SM) error can be reduce below 0.132, then we can have a 5$\sigma$ discovery potential for scalar NP with $\epsilon$ as small as -0.072 with this observable. This observable shows the maximum shift in the position of the scalar NP maxima from the position of SM maxima of all the new observables in this work for a given value of $\epsilon$.
And we have\\
\\
$\frac{\int^{(m_{B}-m_{D})^{2}}_{m_{\tau}^{2}}dq^{2}d[\int^{0}_{-1}d\cos{\theta}(d^{2}\Gamma^{D}_{\tau}/d\cos{\theta}) - \int^{1}_{0}d\cos{\theta}(d^{2}\Gamma^{D}_{\tau}/d\cos{\theta})]/dq^{2}}{\int^{(m_{B}-m_{D})^{2}}_{m_{\tau}^{2}}dq^{2}[d\Gamma(\lambda = +1/2) /dq^{2} - d\Gamma(\lambda = -1/2) /dq^{2}]}|_{SM}$
\be
\begin{split}
- \frac{\int^{(m_{B}-m_{D})^{2}}_{m_{\tau}^{2}}dq^{2}d[\int^{0}_{-1}d\cos{\theta}(d^{2}\Gamma^{D}_{\tau}/d\cos{\theta}) - \int^{1}_{0}d\cos{\theta}(d^{2}\Gamma^{D}_{\tau}/d\cos{\theta})]/dq^{2}}{\int^{(m_{B}-m_{D})^{2}}_{m_{\tau}^{2}}dq^{2}[d\Gamma(\lambda = +1/2) /dq^{2} - d\Gamma(\lambda = -1/2) /dq^{2}]}|_{NP} = 0.221,
\end{split}
\ee
so there is a difference of 0.221 between the $q^{2}$ integrated value of observable $Y_{1}$ in SM compared to the $q^{2}$ integrated value of observable $Y_{1}$ in scalar NP. This means that in the $q^{2}$ integrated value of $Y_{1}$, we only require SM plus Experimental combine error to reduce below 0.044 to have a 5$\sigma$ discovery potential of scalar NP with $\epsilon^{\tau}$ as small as -0.072.

\subsection{$Y_{2}(q^{2}) = \frac{d\Gamma(B \rightarrow D^{*}\tau\nu_{\tau})}{d\Gamma_{D}(\lambda_{\tau}=+1/2) - d\Gamma_{D}(\lambda_{\tau}=-1/2)}$}

\begin{figure}[h!]
\begin{minipage}[t]{0.48\textwidth}
\hspace{0.4cm}
\includegraphics[width=0.9\linewidth, height=6cm]{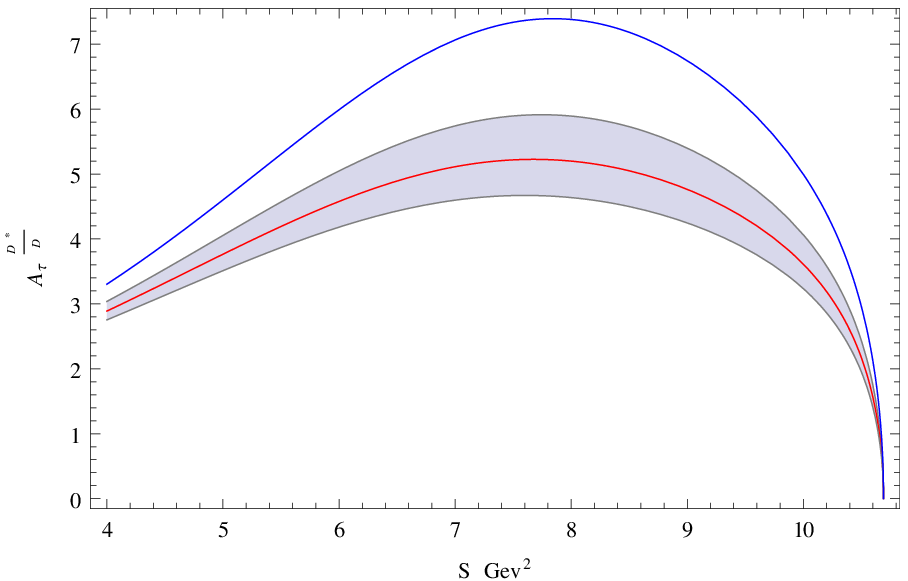}
\end{minipage}
\begin{minipage}[t]{0.48\textwidth}
\hspace{0.4cm}
\includegraphics[width=1\linewidth, height=6cm]{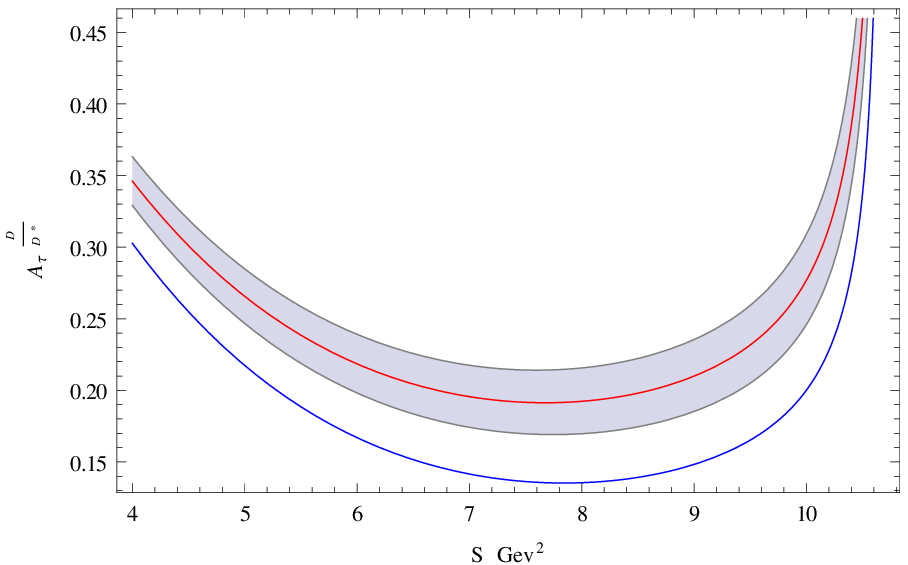}
\end{minipage}
\caption{This figure shows the plots of $Y_{2}(q^{2})$ (left) and $\frac{1}{Y_{2}(q^{2})}$ (Right) where the Blue (SM) and the Red (NP) with the 1 $\sigma$ error bar shown around the NP plot. Here $\epsilon^{\tau}_{s} = \epsilon_{p}^{\tau} = -0.072 \pm 0.028$ is used from Eqs.(\ref{Eq:epsilon}).}
\label{Fig:fig3}
\end{figure}

We can also define another sensitive observable to charged scalar NP as
\be
\begin{split}
Y_{2}(q^{2}) = \frac{d\Gamma(B \rightarrow D^{*}\tau\nu_{\tau})}{d\Gamma_{D}(\lambda_{\tau}=+1/2) - d\Gamma_{D}(\lambda_{\tau}=-1/2)} = \frac{|p_{D^{*}}|}{|p_{D}|}\frac{(|H_{--}|^{2} + |H_{++}|^{2} + |H_{00}|^{2})(1+\frac{m_{\tau}^{2}}{2q^{2}}) + \frac{3m^{2}_{\tau}}{2q^{2}}|\bar{H}_{0t}|^{2}}{|\bar{H}_{t}|^{2}\frac{3m^{2}_{\tau}}{2q^{2}} - (1-\frac{m^{2}_{\tau}}{2q^{2}})|H_{0}|^{2} }
\end{split}
\ee
and a plot of the the observable $Y_{2}(q^{2})$ is shown in the Figure \ref{Fig:fig3}. One of the key feature of this observable is the gap between the SM maxima and NP maxima, which is 2.167. In this observable, the contrast between SM and scalar NP comes out more prominently than any of the other new observables for $\epsilon$ as small as -0.072. But in this observable the shift between the position of SM maxima and scalar NP maxima is very small, about 0.165. $Y_{2}(q^{2})$ is very suitable to test especially models where both $\epsilon_{s}$ and $\epsilon_{p}$ gets substantial contribution.\\
And we have
\be
\frac{\int^{(m_{B}-m_{D^{*}})^{2}}_{m_{\tau}^{2}}dq^{2}d\Gamma(D^{*})/dq^{2}}{\int^{(m_{B}-m_{D})^{2}}_{m_{\tau}^{2}}dq^{2}[d\Gamma(\lambda=+1/2)/dq^{2}-d\Gamma(\lambda=-1/2)/dq^{2}]}|_{NP} = 3.933
\ee
where as
\be
\frac{\int^{(m_{B}-m_{D^{*}})^{2}}_{m_{\tau}^{2}}dq^{2}d\Gamma(D^{*})/dq^{2}}{\int^{(m_{B}-m_{D})^{2}}_{m_{\tau}^{2}}dq^{2}[d\Gamma(\lambda=+1/2)/dq^{2}-d\Gamma(\lambda=-1/2)/dq^{2}]}|_{SM} = 5.292,
\ee
so there is a difference of 1.359 between the $q^{2}$ integrated value of observable $Y_{2}$ in SM compared to the $q^{2}$ integrated value of observable $Y_{2}$ in scalar NP\footnote{this value may depends on the form factors being used for D and $D^{*}$.}. This implies that in $q^{2}$ integrated value of $Y_{2}$, we only require SM plus Experimental combine error to reduce below 0.698 to have a 5$\sigma$ discovery potential with $\epsilon^{\tau}$ as small as -0.072.

\subsection{$Y_{3}(q^{2}) = (\frac{q^{2}}{m^{2}_{\tau}})(A^{D}_{\lambda} + 1)\frac{1}{A^{D}_{\lambda}}$}
Yet another sensitive observable to charged scalar NP can be defined as
\begin{figure}[h!]
\begin{minipage}[t]{0.48\textwidth}
\hspace{0.4cm}
\includegraphics[width=0.9\linewidth, height=6cm]{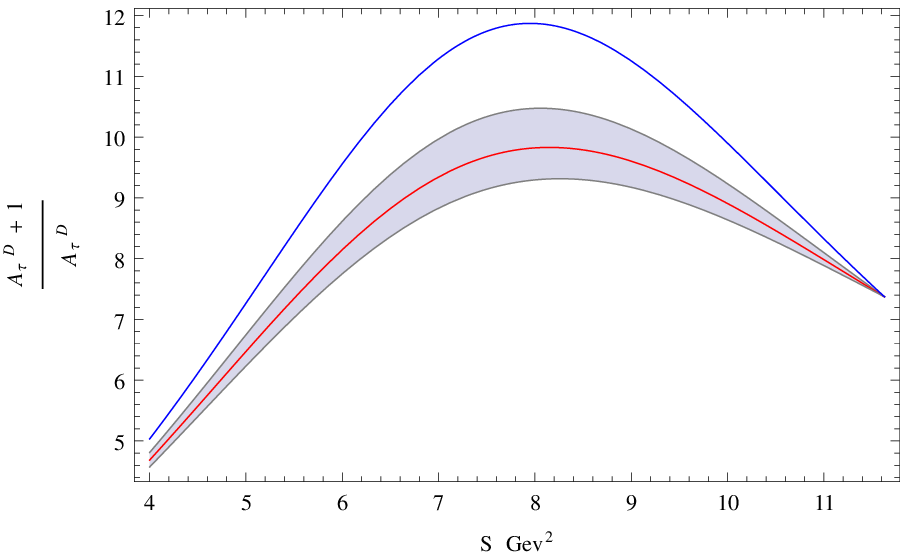}
\end{minipage}
\begin{minipage}[t]{0.48\textwidth}
\hspace{0.4cm}
\includegraphics[width=1\linewidth, height=6cm]{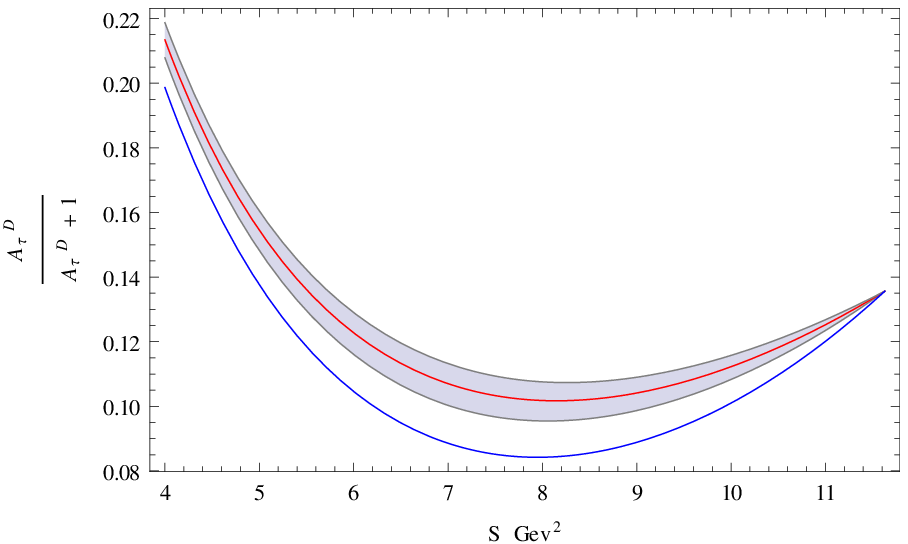}
\end{minipage}
\caption{This figure shows the plots of $Y_{3}(q^{2})$ (left) and $\frac{1}{Y_{3}(q^{2})}$ (Right) where the Blue (SM) and the Red (NP) with the 1 $\sigma$ error bar shown around the NP plot. Here $\epsilon^{\tau}_{s} = \epsilon^{\tau}_{p} = -0.072 \pm 0.028$ is used from Eqs.(\ref{Eq:epsilon}).}
\label{Fig:fig4}
\end{figure}
\be
Y_{3}(q^{2}) = (\frac{q^{2}}{m^{2}_{\tau}})(A^{D}_{\lambda} + 1)\frac{1}{A^{D}_{\lambda}} = \frac{3|\bar{H}_{t}|^{2} + |H_{0}|^{2}}{|\bar{H}_{t}|^{2}\frac{3m^{2}_{\tau}}{2q^{2}} - (1-\frac{m^{2}_{\tau}}{2q^{2}})|H_{0}|^{2}}
\ee
and the plot of this new observable is shown in Figure \ref{Fig:fig4}. The SM (Blue) maxima occurs at $q^{2} = S = 7.948$ with the maxima value of 11.871 where as the scalar NP (Red) maxima occurs at $q^{2} = S = 8.149$ with the maxima value of 9.830, the difference between the positions of the two maxima is 0.201. So if the experimental error in measurements of the position of the maxima and the error in the prediction of the position of the SM maxima can be reduced such that the combined experimental error and theoretical (SM) error can be reduce below 0.041, then we can have a 5$\sigma$ discovery potential for scalar NP with $\epsilon$ as small as -0.072 in this observable. This observable have similar behavior as the observable $\frac{1}{A^{D}_{\lambda}}$ in terms of the shape of the graph as can be seen from comparing Figure \ref{Fig:fig1} and Figure \ref{Fig:fig4}. As the two Figures clearly shows, $Y_{3}$ is more sensitive to the lower $q^{2}$ values than the $\frac{1}{A_{\lambda}^{D}}$, which is more sensitive towards higher $q^{2}$ values. Another big difference between $Y_{3}$ and $\frac{1}{A_{\lambda}^{D}}$ is in their $q^{2}$ integrated values where\\

$\frac{\int_{m_{\tau}^{2}}^{(m_{B}-m_{D})^{2}}dq^{2}\frac{q^{2}}{m_{\tau}^{2}}[2d\Gamma(\lambda = +1/2)]/dq^{2}}{\int_{m_{\tau}^{2}}^{(m_{B}-m_{D})^{2}}dq^{2}[d\Gamma(\lambda = +1/2) /dq^{2} - d\Gamma(\lambda = -1/2]) /dq^{2}]}|_{SM}$
\be
\begin{split}
- \frac{\int_{m_{\tau}^{2}}^{(m_{B}-m_{D})^{2}}dq^{2}\frac{q^{2}}{m_{\tau}^{2}}[2d\Gamma(\lambda = +1/2)]/dq^{2}}{\int_{m_{\tau}^{2}}^{(m_{B}-m_{D})^{2}}dq^{2}[d\Gamma(\lambda = +1/2) /dq^{2} - d\Gamma(\lambda = -1/2]) /dq^{2}]}|_{NP} = 1.216
\end{split}
\label{Eq:y3-dif}
\ee
Comparing the $q^{2}$ integrated value of about 1.216 for $Y_{3}$ above to the $q^{2}$ integrated value of about 0.551 for the $\frac{1}{A_{\lambda}^{D}}$, it is clear that the observable $Y_{3}$ is much more sensitive observable to charged scalar than $\frac{1}{A_{\lambda}^{D}}$. Also from Eqs.(\ref{Eq:y3-dif}) we see that for observable $Y_{3}$, we only require SM plus Experimental combine error to reduce below 0.243 to have a 5$\sigma$ discovery potential of scalar NP with $\epsilon^{\tau}$ as small as -0.072. Similar observable can be defined from $A_{\lambda}^{D^{*}}$, however this observable is not that sensitive to charged scalars due to supression of charged scalar coupling in $D^{*}$ final state by a factor of $\frac{m_{b}-m_{c}}{m_{b}+m_{c}}$ relative to D final state in $B \rightarrow D^{(*)} \tau \nu_{\tau}$.

\section{Conclusions.}

In this work we have given four new observables which are very sensitive to the presence of charged scalars in the $B \rightarrow D^{(*)}\tau \nu_{\tau}$ decays. All the new observables shows substantial deviation from SM values in two main features of them i.e (1) in presence of charged scalar, they all show substantial shift in the position of the maxima from that of the SM value and (2) in presence of charged scalar, they also show substantial deviation in their $q^{2}$ integrated value from that of the SM one. In the following we will enumerate the key results for each new observables from the preceding analysis.
\begin{enumerate}
\item $\frac{1}{A_{\lambda}^{D}}$
\begin{enumerate}
\item The shift in the position of the maxima due to the presence of the charged scalar from SM in this observable turn out to be 0.387. This implies that to have a 5 $\sigma$ discovery potential of charged scalar with $\epsilon^{\tau}$ as small as -0.072, we only require the combine theoretical (SM) error and experimental error in the measurement of the position of this maxima to reduce just below 0.078.
\item The difference in the $q^{2}$ integrated value of $\frac{1}{A_{\lambda}^{D}}|_{SM}$ and $\frac{1}{A_{\lambda}^{D}}|_{NP}$ turn out to be 0.551, so we only need to reduce the combine theoretical (SM) and experimental errors in the measurement of the $q^{2}$ integrated value of $\frac{1}{A_{\lambda}^{D}}$ just below 0.110 to have a 5 $\sigma$ discovery potential of charged scalar with $\epsilon^{\tau}$ as small as -0.072.
\item In the $q^{2}$ integrated value, $\frac{1}{A_{\lambda}^{D}}$ is a better observable than $A_{\lambda}^{D}$, in probing the presence of charged scalar, by about an order of magnitude.
\end{enumerate}
\item $Y_{1}(q^{2}) = \frac{A^{D}_{\theta}}{A^{D}_{\lambda}} = \frac{\frac{3m^{2}_{\tau}}{2q^{2}}Re(H_{0}\bar{H}^{*}_{t})}{\frac{3m_{\tau}^{2}}{2q^{2}}|\bar{H}_{t}|^{2} - (1 - \frac{m_{\tau}^{2}}{2q^{2}})|H_{0}|^{2}}.$
\begin{enumerate}
\item The shift in the position of the maxima due to the presence of the charged scalar from SM in this observable turn out to be 0.659. This implies that to have a 5 $\sigma$ discovery potential of charged scalar with $\epsilon^{\tau}$ as small as -0.072, we only require the combine theoretical (SM) error and experimental error in the measurement of the position of this maxima to reduce just below 0.132.
\item The difference in the $q^{2}$ integrated value of $Y_{1}(q^{2})|_{SM}$ and $Y_{1}(q^{2})|_{NP}$ turn out to be 0.221, so we only need to reduce the combine theoretical (SM) error and experimental error in the measurement of the $q^{2}$ integrated value of $Y_{1}(q^{2})$ just below 0.044 to have a 5 $\sigma$ discovery potential of charged scalar with $\epsilon^{\tau}$ as small as -0.072.
\end{enumerate}
\item $Y_{2}(q^{2}) = \frac{d\Gamma(B \rightarrow D^{*}\tau\nu_{\tau})}{d\Gamma_{D}(\lambda_{\tau}=+1/2) - d\Gamma_{D}(\lambda_{\tau}=-1/2)} = \frac{|p_{D^{*}}|}{|p_{D}|}\frac{(|H_{--}|^{2} + |H_{++}|^{2} + |H_{00}|^{2})(1+\frac{m_{\tau}^{2}}{2q^{2}}) + \frac{3m^{2}_{\tau}}{2q^{2}}|\bar{H}_{0t}|^{2}}{|\bar{H}_{t}|^{2}\frac{3m^{2}_{\tau}}{2q^{2}} - (1-\frac{m^{2}_{\tau}}{2q^{2}})|H_{0}|^{2} }$
\begin{enumerate}
\item The shift in the position of the maxima due to the presence of the charged scalar from SM in this observable turn out to be 0.165. This implies that to have a 5 $\sigma$ discovery potential of charged scalar with $\epsilon^{\tau}$ as small as -0.072, we only require the combine theoretical (SM) error and experimental error in the measurement of the position of this maxima to reduce just below 0.033.
\item The difference in the $q^{2}$ integrated value of $Y_{2}(q^{2})|_{SM}$ and $Y_{2}(q^{2})|_{NP}$ turn out to be 1.359, so we only need to reduce the combine theoretical (SM) error and experimental error in the measurement of the $q^{2}$ integrated value of $Y_{2}(q^{2})$ just below 0.698 to have a 5 $\sigma$ discovery potential of charged scalar with $\epsilon^{\tau}$ as small as -0.072.
\item (Note$^{\dagger}$) Results in this observable may depends on the different form factors for D and $D^{*}$  being used.
\end{enumerate}
\item $Y_{3}(q^{2}) = (\frac{q^{2}}{m^{2}_{\tau}})(A^{D}_{\lambda} + 1)\frac{1}{A^{D}_{\lambda}} = \frac{3|\bar{H}_{t}|^{2} + |H_{0}|^{2}}{|\bar{H}_{t}|^{2}\frac{3m^{2}_{\tau}}{2q^{2}} - (1-\frac{m^{2}_{\tau}}{2q^{2}})|H_{0}|^{2}}$
\begin{enumerate}
\item The shift in the position of the maxima due to the presence of the charged scalar from SM in this observable turn out to be 0.201. This implies that to have a 5 $\sigma$ discovery potential of charged scalar with $\epsilon^{\tau}$ as small as -0.072, we only require the combine theoretical (SM) error and experimental error in the measurement of the position of this maxima to reduce just below 0.041.
\item The difference in the $q^{2}$ integrated value of $Y_{3}(q^{2})|_{SM}$ and $Y_{3}(q^{2})|_{NP}$ turn out to be 1.216, so we only need to reduce the combine theoretical (SM) error and experimental error in the measurement of the $q^{2}$ integrated value of $Y_{3}(q^{2})$ just below 0.243 to have a 5 $\sigma$ discovery potential of charged scalar with $\epsilon^{\tau}$ as small as -0.072.
\end{enumerate}

\end{enumerate}
So in short, we have proposed four new observables which are very sensitive towards presence of new charged scalars. Some of these new observables are more sensitive then observables such as $R(D^{(*)})$ and $A_{\lambda}^{D^{(*)}}$ by an order of magnitude. Most of these observables are related to D mode final sates, we can define similar observables in the $D^{*}$ mode final states but even though these new observables are expected to be much more sensitive then observables such $R(D^{*})$ and $A_{\lambda}^{D^{*}}$, the sensitivities of $D^{*}$ mode final observables are supressed by a factor of $\frac{m_{b}-m_{c}}{m_{b}+m_{c}}$ relative to D mode final states.

{\large Acknowledgments: \large} This work is supported and funded by the Department of Atomic Energy of the Government of India and by the Government of Tamil Nadu.

\end{document}